\documentclass[aps,pra,twocolumn,showpacs,superscriptaddress]{revtex4}

\usepackage{dcolumn}
\usepackage[latin1]{inputenc}
\usepackage[english]{babel}
\usepackage[dvips]{graphicx}
\usepackage{enumerate,amsthm,amsmath,amssymb,color,graphicx,bbm,float}
\usepackage{bm}
\usepackage{mathtools}
\usepackage{natbib}
\usepackage{subfigure}
\usepackage{synttree}

\usepackage{epstopdf}

\usepackage{tikz}
\usetikzlibrary{arrows}

\newcommand{\be}{\begin{equation}}
\newcommand{\ee}{\end{equation}}

\def\multiset#1#2{\ensuremath{\left(\kern-.3em\left(\genfrac{}{}{0pt}{}{#1}{#2}\right)\kern-.3em\right)}}

\begin{document}

\preprint{APS/123-QED}

\title{Preventing Calibration Attacks on the Local Oscillator\\in Continuous-Variable Quantum Key Distribution}

\author{Paul Jouguet}\affiliation{LTCI, CNRS - Telecom ParisTech, 46 rue Barrault, 75013 Paris, France}\affiliation{SeQureNet, 23 avenue d'Italie, 75013 Paris, France}
\author{S\'ebastien Kunz-Jacques}\affiliation{SeQureNet, 23 avenue d'Italie, 75013 Paris, France}
\author{Eleni Diamanti}\affiliation{LTCI, CNRS - Telecom ParisTech, 46 rue Barrault, 75013 Paris, France}

\date{\today}

\begin{abstract}
Establishing an information-theoretic secret key between two parties using a quantum key distribution (QKD) system is only possible when an accurate characterization of the quantum channel and proper device calibration routines are combined. Indeed, security loopholes due to inappropriate calibration routines have been shown for discrete-variable QKD. Here, we propose and provide experimental evidence of an attack targeting the local oscillator calibration routine of a continuous-variable QKD system. The attack consists in manipulating the classical local oscillator pulses during the QKD run in order to modify the clock pulses used at the detection stage. This allows the eavesdropper to bias the shot noise estimation usually performed using a calibrated relationship. This loophole can be used to perform successfully an intercept-resend attack. We characterize the loophole and suggest possible countermeasures.
\end{abstract}

\maketitle

\section{Introduction}
The two communicating parties of a quantum key distribution (QKD) protocol \cite{SBC08}, Alice and Bob, can in principle share an information-theoretic secret key after the exchange of a large number of quantum signals through a physical channel, known as quantum channel, which is subject to eavesdropping, and additional information sent on a public but authenticated classical channel. After Alice and Bob have agreed on a set of non-commuting quantum operators, they can safely encode the key into these variables: any eavesdropping attempt disturbs the transmitted quantum states and is discovered after random sampling of a fraction of Alice and Bob's correlated data. However, deviations of the practical implementation of a QKD protocol from the underlying theoretical model can be exploited by an eavesdropper.

In most commonly used QKD systems, the key information is encoded on discrete variables, such as the polarization of a single photon, and thus specific components for single-photon detection are required. Exploiting imperfections of such devices has led to powerful attacks, namely the time-shift attack \cite{ZFQ+08}, the phase-remapping attack \cite{XQL10}, and the remote control of single-photon detectors using tailored bright illumination \cite{LWW10}. Other attacks proposed against discrete-variable QKD systems include Trojan horse \cite{GFK+:pra06}, device calibration \cite{JWL11}, and wavelength dependent beamsplitter \cite{LWH+:pra11} attacks. The latter have also been adapted to continuous-variable QKD (CVQKD), where the key information is encoded on continuous variables \cite{WPG:rmp12}, such as the quadratures of coherent states \cite{GG02}. In CVQKD systems, measurements are performed using standard coherent detection techniques, in particular homodyne detection when the protocol requires the measurement of a single quadrature of the electromagnetic field or heterodyne detection when both quadratures need to be measured. Wavelength dependent beamsplitter attacks targeting CVQKD schemes using heterodyne detection have recently been studied \cite{HWY+:arxiv13, MSJL:arxiv13b}. Finally, attacks specific to CVQKD \cite{FGG:iqec07,MSJL:arxiv13a} typically involve manipulation of the power of the local oscillator, which is the phase reference classical signal required for the coherent detection and is usually sent from Alice to Bob together with the quantum signal \cite{JKL+:natphoton13}.

Here, we consider device calibration attacks against continuous-variable QKD. These attacks arise from a subtle link between the local oscillator calibration procedure and the clock generation procedure in practical CVQKD setups using Gaussian modulation of coherent states and homodyne detection. We show that combining this security loophole with intercept-resend attacks can compromise the security of continuous-variable QKD in the absence of appropriate countermeasures. With recent advances in this technology, which allows for long-distance key distribution using standard telecommunication components and with strong security guarantees \cite{JKL+:natphoton13}, assuring the practical security of all aspects of the implementation, and specifically of the ubiquitous calibration procedure, is of great importance.

\section{Security assumptions and calibration techniques}
A standard assumption when designing and implementing a CVQKD system is that the local oscillator cannot be manipulated by an eavesdropper. This cannot, however, be verified in practice since the local oscillator is a classical, and therefore intense, signal, and thus the no-cloning theorem does not apply. This means that the local oscillator can be measured and regenerated, or directly amplified, without adding any additional disturbance.

Current security proofs do not explicitly take into account the local oscillator, which is not required at a theoretical level to define the exchanged states  and the performed measurements \cite{NGA:prl06, GC:prl06, LGR13}. In particular, all the quantities that are used in the calculation of the secret key generation rate are expressed in shot noise units. Knowledge of the shot noise is therefore required. In principle, the shot noise variance can be evaluated using a balanced homodyne detector, as the variance of the interference between the local oscillator and the vacuum mode. This measurement method incurs some statistical uncertainty due to the finite size of the data, as was studied in \cite{JKDL:pra12}. Alternatively, the linear relationship between the variance of this measurement and the input power of the local oscillator signal on the homodyne detector can be used to estimate the shot noise during the quantum transmission, provided that the local oscillator power is known.

\begin{figure}
\centering
 \includegraphics[width=80mm]{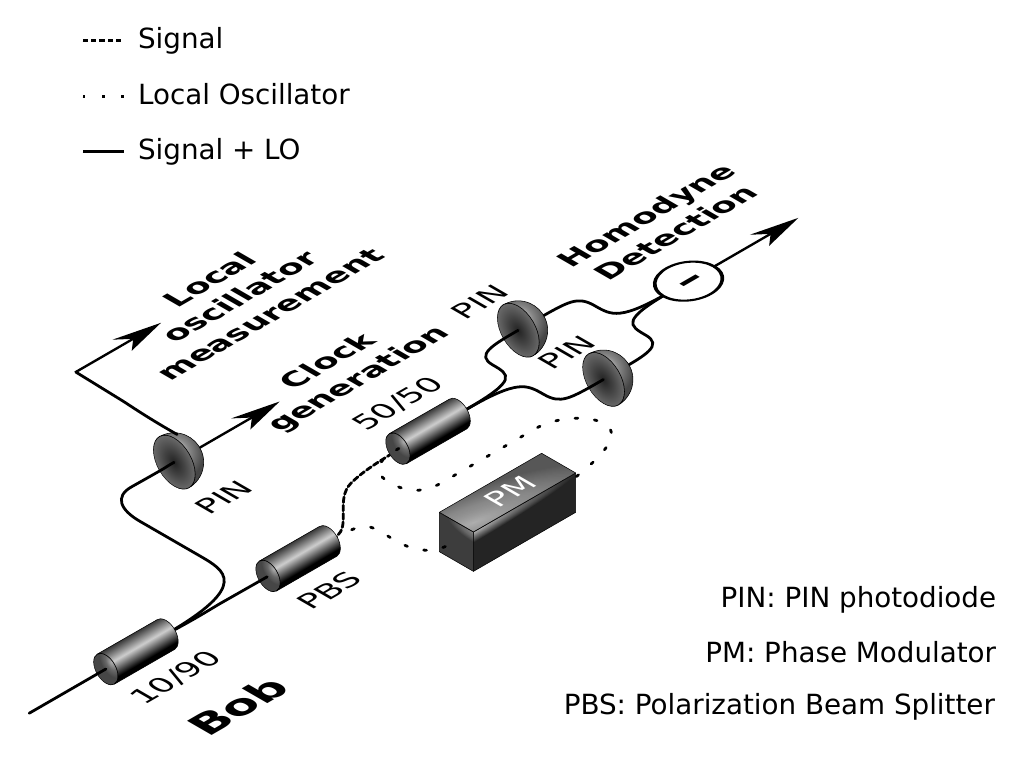}
  \caption{Local oscillator experimental measurement procedure. Here a PIN photodiode at the entrance of Bob's setup is used for two purposes: generating a clock on Bob's side and generating a signal proportional to the local oscillator power.}
   \label{figure:lo_measurement}
\end{figure}

A standard calibration technique, used for instance in  \cite{JKD:oe12}, consists in establishing in a secure laboratory, before the QKD run, the aforementioned linear relationship between the shot noise and the local oscillator power. During the QKD run, the local oscillator power is measured either with a power meter or with a photodiode followed by an integration circuit, at the input of Bob's site. In either case, a signal proportional to the intensity of the local oscillator over a time period that should be equal to the homodyne detection integration window is available. The previously established linear relationship can then be used to deduce the shot noise level used for the secret key rate calculation. This approach, however, has two shortcomings. First, it is not possible to trust the power of the signal entering Bob's device, since an eavesdropper can easily add another classical signal (for instance, at a different wavelength) into the quantum channel. Second, in a practical CVQKD system, the local oscillator is not only used as an intense signal coherent with the weak quantum signal and therefore allowing to measure its quadratures; it is also used to generate the clock signal that is necessary to perform the measurements, as shown in Fig. \ref{figure:lo_measurement}. Therefore, the local oscillator signal can be suitably modified by an eavesdropper such that the trigger signal generated by the clock circuit is also altered.

In the following, we describe how the interplay between the local oscillator calibration and the clock generation procedures can be exploited to perform an eavesdropping attack.

\begin{figure}
     \begin{center}
        \subfigure{
            \label{figure:trigger_profile}
            \includegraphics[width=0.4\textwidth]{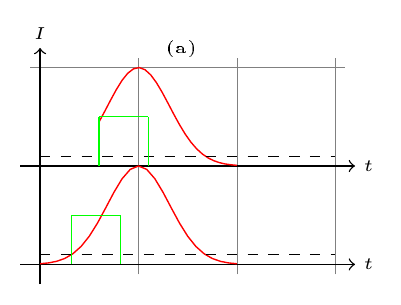}
        }\\ %  ------- End of the first row ----------------------%
        \subfigure{
             \label{figure:hd_profile}
           \includegraphics[width=0.4\textwidth]{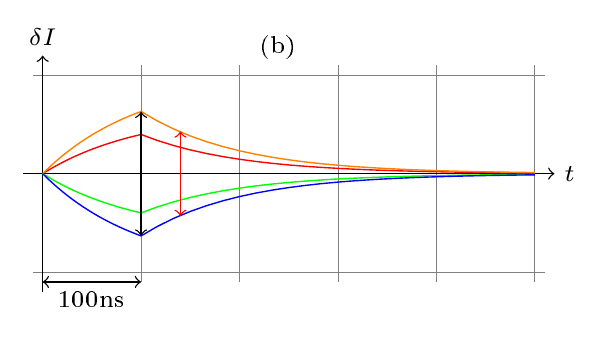}
        }\\
    \end{center}
    \caption{
        (a) Profile of the trigger signal generated at Bob's site depending on the shape of the local oscillator pulse. (b) Differential signal obtained by the homodyne detector for several modulated quadratures. After an integration period of $\Delta = 100$ ns, the capacitor discharges exponentially. Depending on the time of the measurement, the variance of the measurement of the homodyne detection is different.
     }
   \label{fig:triggerHDsubfigures}
\end{figure}

%Modification of the shape of a local oscillator pulse and consequence on the trigger generation and on the statistics of the homodyne detection.

\section{Description of the local oscillator calibration attack}
\begin{figure}
\centering
    \includegraphics[width=0.4\textwidth]{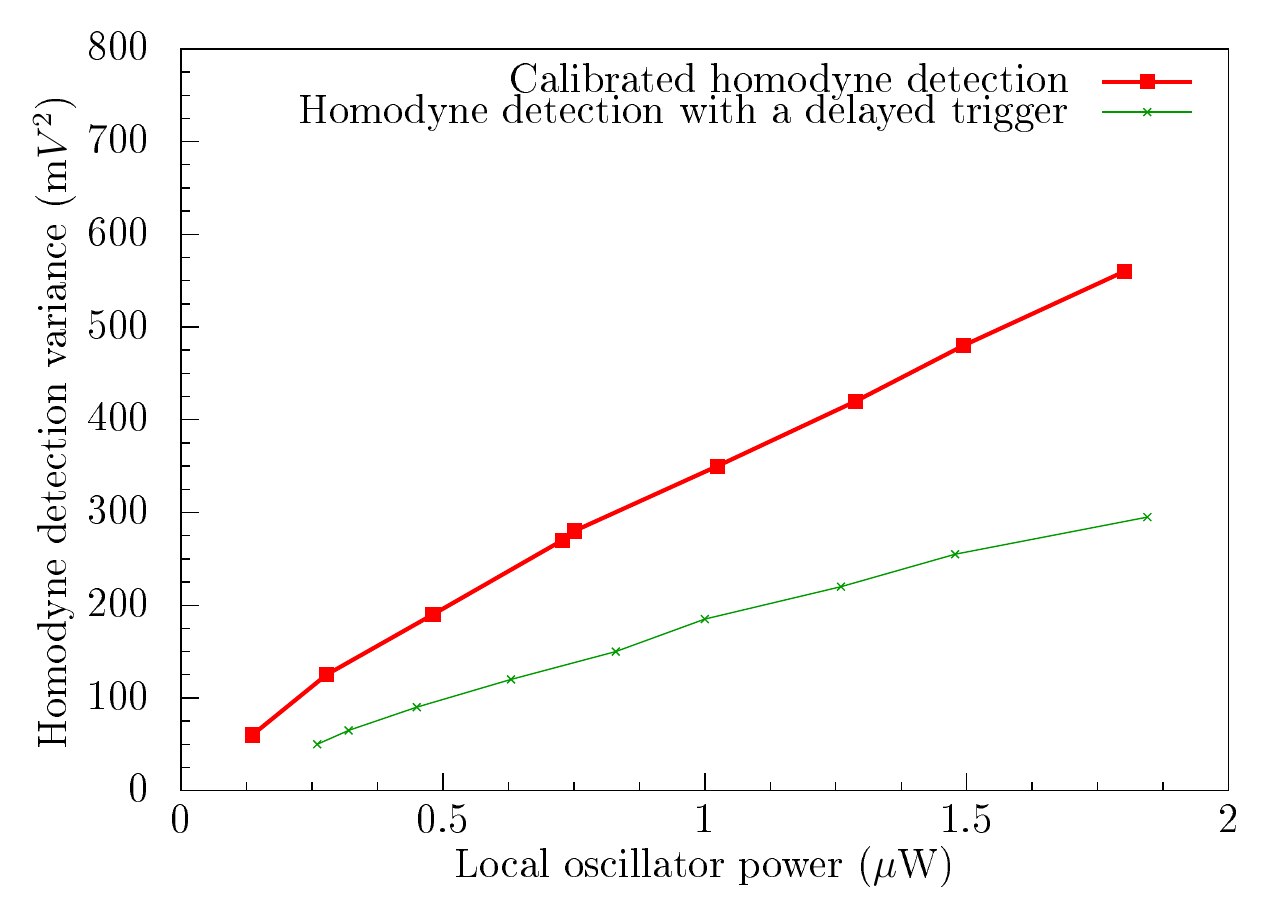}
    \caption{(Color online) In red, the calibrated linear relationship between the variance of the homodyne detection measurements and the local oscillator power. In green, the linear relationship we obtain when delaying the trigger of the homodyne detection by 10 ns.}
     \label{figure:hd_stats}
\end{figure}

The basic principle of the attack is illustrated in Figs. \ref{fig:triggerHDsubfigures} and \ref{figure:hd_stats}. In particular, as shown in Fig. \ref{figure:trigger_profile}, the clock circuit is usually designed to output a rising trigger signal when the intensity entering the photodiode is above a certain threshold. Subsequently, this trigger is delayed such that the value of the signal at the output of the homodyne detection is maximized. A potential attack for an eavesdropper consists in attenuating the beginning of the local oscillator pulse, which induces a delay of the trigger used for the measurements. Note that this was also suggested in \cite{CQZ+:njp11} as a potential source of loophole. Figure \ref{figure:hd_stats} shows experimental results illustrating the relationship between the variance of the measurement on the homodyne detection and the local oscillator power for different trigger signals. These results were obtained using the setup of Fig. \ref{figure:lo_measurement}, which corresponds to a simplified version of Bob's setup employed for long-distance continuous-variable QKD using Gaussian modulation of coherent states \cite{JKL+:natphoton13}. The experiment shows that a delayed trigger results in a decrease of the detection response slope. This is because a homodyne measurement is usually performed by integrating the differential photocurrent during a period $\Delta$ using an integrator circuit: after this period $\Delta$, the capacitor discharges exponentially, which implies that the maximum measurement variance is obtained when the trigger coincides with the end of the period $\Delta$, as shown in Fig. \ref{figure:hd_profile}. As a result, if Alice and Bob use the previously calibrated relationship to evaluate the shot noise based on the measured local oscillator power, they will use a false value, if the trigger signal has been delayed during the QKD run. In particular, they will overestimate the value of the shot noise, and consequently underestimate the excess noise present in the setup. This creates an important loophole in the security of the implementation.

Based on this loophole, we propose the following practical attack. It is important to note that this attack can be implemented with current technology, without any need, for instance, for a quantum memory.
\begin{itemize}
 \item The eavesdropper, Eve, introduces a phase-independent attenuator in the quantum channel and applies an attenuation factor $\alpha$ ($0 \leq \alpha \leq 1$) on a fraction $\nu$ ($0 \leq \nu \leq 1$) of the local oscillator pulses in order to modify their shape. The trigger used to perform the homodyne measurement relative to these pulses is delayed by $\delta$.
 \item Eve introduces a beam splitter in the quantum channel and for a fraction $\mu$ ($0 \leq \mu \leq 1$) of the input signal pulses she measures both quadratures and prepares the appropriate quantum state, whereas for a fraction $1-\mu$ of the input signal pulses she just eavesdrops using the beamsplitter. This so called partial intercept-resend attack was implemented experimentally in \cite{LDG+:prl07}.
\end{itemize}

When Eve increases the fraction $\mu$ of signal pulses over which she performs an intercept-resend attack, she introduces more noise, which lowers the amount of secret key that Alice and Bob can extract from the quantum transmission. The fraction $\nu$ of local oscillator pulses attenuated by Eve and the attenuation factor $\alpha$ are two free parameters that play the same role: they scale the variance of the measurements made by Bob while his shot noise estimation remains unchanged. This leads Alice and Bob to conclude that no noise has been introduced in the quantum channel and hence they establish a key without detecting the presence of Eve.

\section{Analysis of the excess noise}
To assess the impact of our attack on the security of continuous-variable QKD, we detail the parameter estimation procedure that is necessary for the derivation of the secret key and how this procedure is altered when the attack is implemented. In a practical CVQKD setup, Alice and Bob estimate the quantities required to compute the secret key rate by sampling $m=N-n$ couples of correlated variables $(x_i,y_i)_{i=1\dots m}$, where $N$ is the total number of quantum signals sent through the quantum channel and $n$ is the number of signals used for the key establishment. Since for CVQKD it is sufficient to estimate the covariance matrix of the state shared by Alice and Bob, the only parameters that need to be estimated are the variance on Alice's and Bob's sites, $\langle x^2\rangle$ and $\langle y^2\rangle$, respectively, and the covariance between Alice and Bob, namely $\langle xy\rangle$ (assuming here that $x$ and $y$ are centered variables, that is, that $\langle x\rangle = \langle y \rangle =0$). Then, the following estimators are used during the QKD run:
\begin{align}
 \langle x^2\rangle &= V_A, \langle xy\rangle = \sqrt{\eta T}V_A\\
 \langle y^2\rangle &= \eta T V_A + N_0 + \eta T \xi + v_\mathrm{el}
\end{align}
%while during the calibration procedure the relevant estimator is:
%\begin{align}
% \langle {y_0}^2\rangle &= N^{'}_0 + v_\mathrm{el}.
%\end{align}
In the above expressions, $T$ is the quantum channel transmittance, $V_A$ is the modulation variance, $\xi$ is the excess noise, $N_0$ is the shot noise,
% during the QKD run, $N^{'}_0$ is the shot noise measured during the calibration procedure,
$\eta$ is the efficiency of the homodyne detector, and $v_\mathrm{el}$ is the electronic noise (all expressed in their respective units).

Here we assume that the electronic noise does not change between the QKD run and the calibration procedure. In theory, an eavesdropper may also try to modify the value of the electronic noise, for example by changing the temperature operating conditions of the electronic circuit of the homodyne detection between the calibration and the QKD run. However, the impact of such an attack would be less significant since the value of the electronic noise is typically between 10 and 20 dB below the shot noise.

In order to compute confidence intervals for these parameters, we consider a normal model for Alice and Bob's correlated variables $(x_i,y_i)_{i=1\dots m}$, namely $y=tx+z$, where $t=\sqrt{\eta T}\in \mathbb{R}$, and $z$ follows a centered normal distribution with unknown variance $\sigma^2=N_0+\eta T\xi+v_\mathrm{el}$. Note that this normal model is an assumption justified in practice but not by current proof techniques, which show that the Gaussian assumption is valid once the covariance matrix is known \cite{LGR13}.

Maximum-Likelihood estimators $\hat{t}$, $\hat{\sigma}^2$ and $\hat{V}_A$ are known for the normal linear model:
\begin{equation}
 \hat{t}=\frac{\sum_{i=1}^m x_i y_i}{\sum_{i=1}^m x_i^2}, \hat{\sigma}^2=\frac{1}{m}\sum_{i=1}^m(y_i-\hat{t}x_i)^2, \hat{V}_A=\frac{1}{m}\sum_{i=1}^m {x}_i^2\nonumber
\end{equation}
These are independent estimators with distributions:
\begin{equation}
\hat{t} \sim \mathcal{N}\left(t,\frac{\sigma^2}{\sum_{i=1}^m x_i^2}\right),\frac{m\hat{\sigma}^2}{\sigma^2}, \frac{m\hat{V}_A}{V_A} \sim \chi^2(m-1),\nonumber
\end{equation}
where $t$, $\sigma^2$ and $V_A$ are the true values of the parameters. Using the previous estimators and their confidence intervals together with the shot noise value from the calibration $N'_0$, it is then possible to estimate $T=\hat{t}^2/\eta$ and $\xi=(\hat{\sigma}^2-N'_0-v_\mathrm{el})/\hat{t}^2$.

If the eavesdropper can change the slope of the homodyne detection response as previously explained, the equality $N'_0 = N_0$ is not verified. This leads to the following estimation for the excess noise when a calibration attack occurs:
\begin{equation}
 \hat{\xi}_{\text{calib}}= \hat{\xi} + \frac{N^{'}_0-N_0}{\hat{t}^2},
\end{equation}
where $\hat{\xi}$ is the estimate without the attack. In order to compute a secret key rate, the excess noise must be expressed in shot noise units, hence we have:
\begin{equation}
 \frac{\hat{\xi}_{\text{calib}}}{N'_0}= \frac{N_0}{N'_0} \left[\frac{\hat{\xi}}{N_0} + \frac{1}{\hat{t}^2}\left(1-\frac{N^{'}_0}{N_0}\right) \right]\label{eq:xicalib}
\end{equation}

Next, we consider the excess noise introduced by a partial intercept-resend (PIR) attack alone. According to the analysis of \cite{LDG+:prl07}, in this case, the probability distribution of Bob's measurements is the weighted sum of two Gaussian distributions with a weight of $\mu$ for the intercepted and resent data and a weight of $1-\mu$ for the transmitted data:
\begin{align}
 \langle y^2\rangle_{\text{IR}} &= \eta T (V_A+2N_0) + N_0 + \eta T \xi + v_\mathrm{el}\\
 \langle y^2\rangle_{\text{BS}} &= \eta T V_A + N_0 + \eta T \xi + v_\mathrm{el},
\end{align}
where $\xi$ is the technical excess noise of the system. The excess noise introduced by this attack can then be computed as:
\begin{equation}
 \hat{\xi}^{\text{PIR}}=\hat{\xi}+2\mu N_0 \label{eq:xiPIR}
\end{equation}
In practice, when a full intercept-resend attack is implemented ($\mu=1$), the excess noise is dominated by the second term in the above expression due to the noise introduced by Eve's measurements.

If, additionally, the eavesdropper performs the local oscillator calibration attack, then the excess noise introduced by the partial intercept-resend attack is computed, in shot noise units, as:
\begin{align}
 \frac{\hat{\xi}^{\text{PIR}}_{\text{calib}}}{N'_0}= \frac{N_0}{N'_0} \left[ \frac{\hat{\xi}^{\text{PIR}}}{N_0} + \frac{1}{\hat{t}^2}\left (1-\frac{N^{'}_0}{N_0} \right) \right]\label{eq:xiPIRcalib}
\end{align}

\section{A quantitative example}
When the eavesdropper implements a full intercept-resend attack ($\mu=1$), and with a typical value of $\xi/N_0 = 0.1$, we find from Eq. (\ref{eq:xiPIR}) that the noise introduced by the attack is $\xi^{\text{PIR}}/N_0 = 2.1$. This noise value is above the entanglement breaking limit, hence no secret key can be exchanged, independently of the communication distance. However, if Eve implements additionally the local oscillator calibration attack, then Alice and Bob will estimate the excess noise using Eq. (\ref{eq:xiPIRcalib}). For example, for a transmission $T=0.5$ and a homodyne detection efficiency $\eta=0.5$, we find:
\begin{align}
 \frac{\hat{\xi}^{\text{PIR}}_{\text{calib}}}{N'_0}= \frac{N_0}{N'_0} \left[2.1+\frac{1}{0.5\times0.5}\left(1-\frac{N^{'}_0}{N_0}\right)\right]
\end{align}
Then, for $N^{'}_0/N_0 \approx 1.5$, which is a realistic value as shown in Fig. \ref{figure:hd_stats}, the excess noise estimated by Alice and Bob will be close to zero, hence they will conclude they can share a secret key. The security of the protocol is thus entirely compromised.

\section{Countermeasure: real-time shot noise measurement techniques}
\begin{figure}[ht!]
     \begin{center}
        \subfigure[Real-time shot noise measurement using an amplitude modulator on Bob's signal path.]{
            \label{figure:shot_noise_measurement_with_attenuator}
            \includegraphics[width=0.4\textwidth]{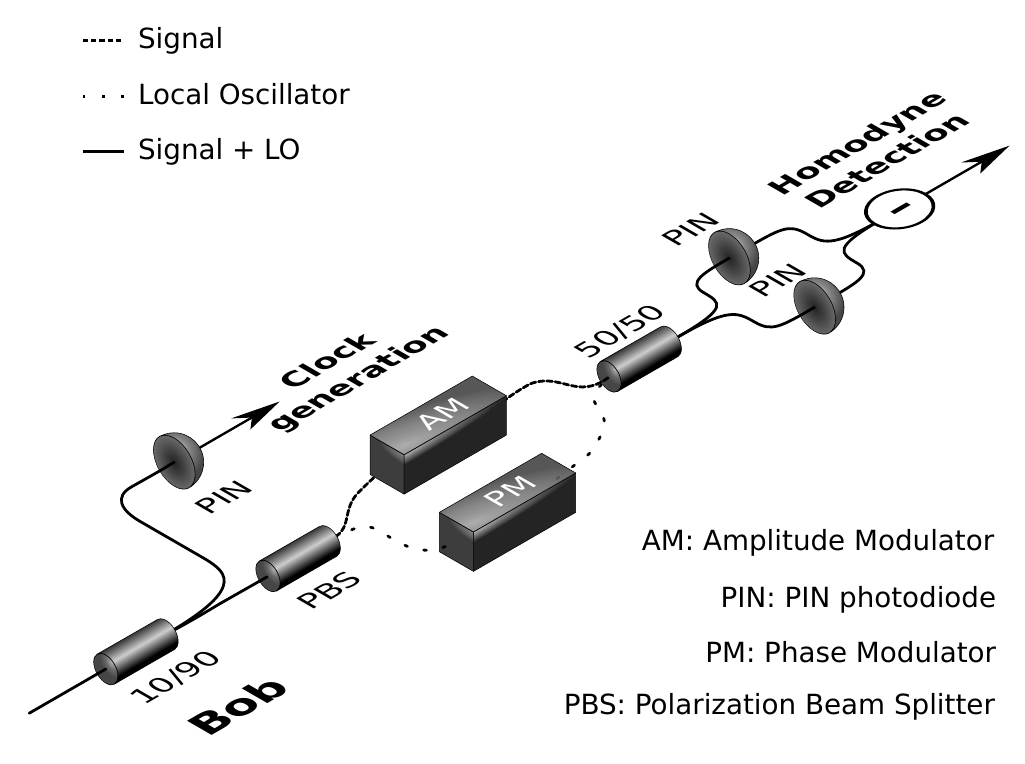}
        }\\ %  ------- End of the first row ----------------------%
        \subfigure[Real-time shot noise measurement using a second homodyne detection on Bob's local oscillator path.]{
           \label{figure:shot_noise_measurement_with_other_HD}
           \includegraphics[width=0.4\textwidth]{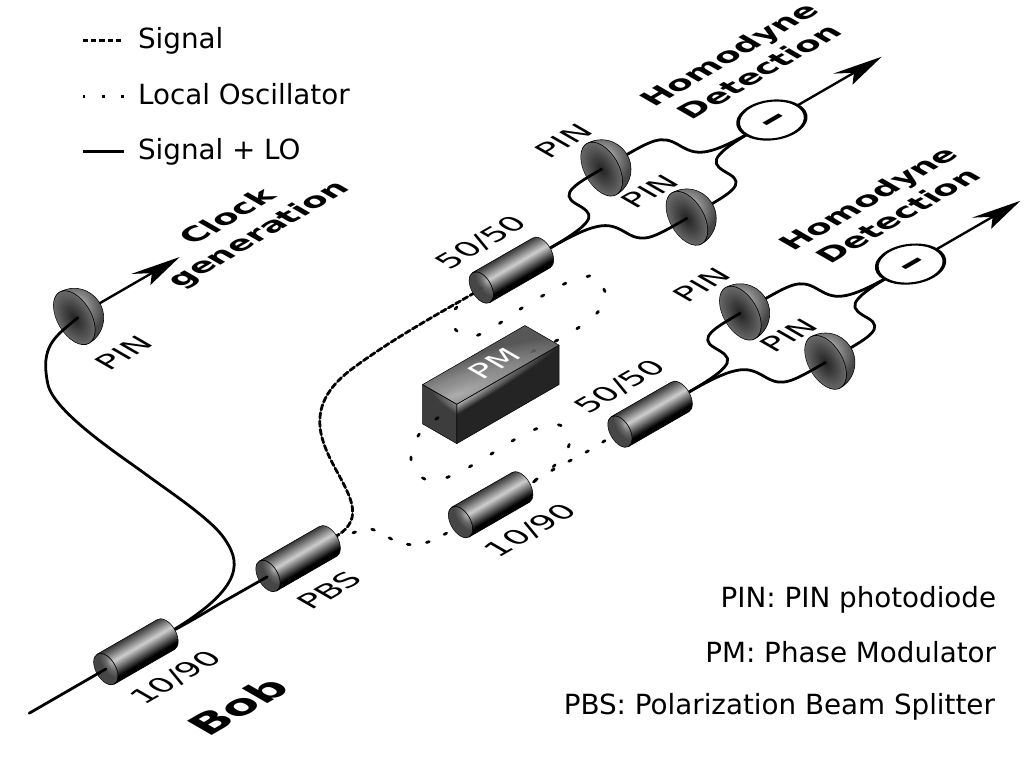}
        }\\
    \end{center}
    \caption{
        Real-time shot noise measurement procedures protecting a CVQKD system against a local oscillator calibration attack.
     }
   \label{fig:shot_noise_measurement_subfigures}
\end{figure}

In practice, it is possible to show that a calibrated linear relationship between the shot noise level and local oscillator power cannot be used in the presence of an eavesdropper (see Appendix for a detailed analysis). Therefore, a countermeasure for the proposed attack consists in devising techniques allowing to measure the shot noise in real time. One such technique consists in applying a strong attenuation on Bob's signal path to a randomly chosen set of pulses, using, for instance, an optical switch or an amplitude modulator. Alternatively, an additional homodyne detector dedicated to the real-time shot noise measurement can be used: a beam splitter is introduced in Bob's local oscillator path and the relative sensitivity of the two homodyne detectors is calibrated. A schematic representation of the two techniques is shown in Fig. \ref{fig:shot_noise_measurement_subfigures}. In both methods, two noise measurements on two sets of pulses alow to extract the shot noise and the signal noise by inverting a linear system. To the best of our knowledge, none of these techniques has been proposed or implemented in CVQKD.

\begin{figure}
\centering
    \includegraphics[width=0.4\textwidth]{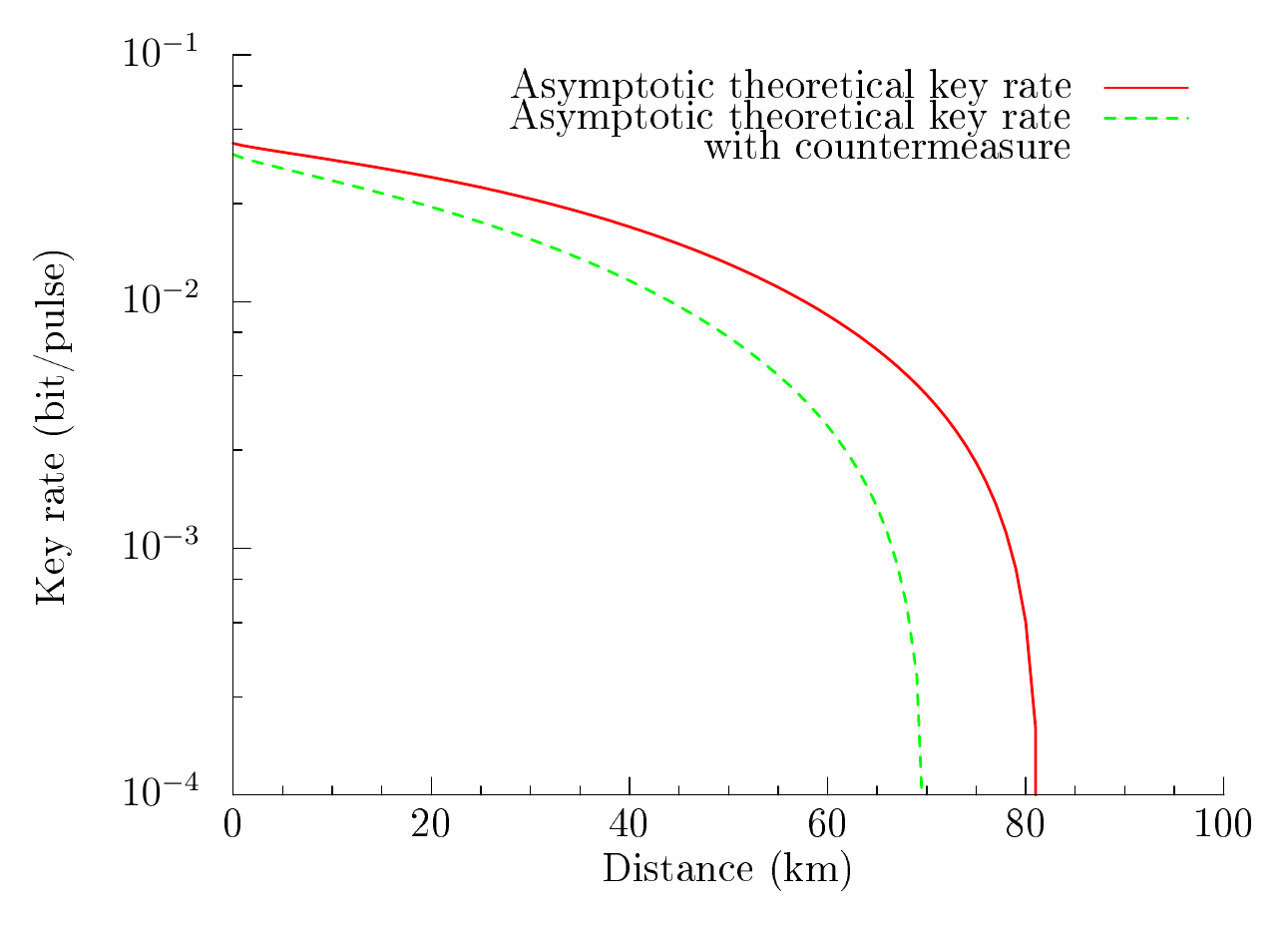}
    \caption{(Color online) Secret key rate for collective attacks in the asymptotic regime. Both plots are obtained in the so-called realistic model where the electronic noise and the efficiency of the homodyne detection are calibrated and cannot be altered by the eavesdropper. The upper plot corresponds to the secret key rate computed without implementing any countermeasure against the local oscillator calibration attack. The lower plot is obtained when inserting an optical switch with typical losses of 2.7 dB on Bob's signal path and discarding 10\% of the pulses on Bob's side at random to perform a real-time shot noise measurement. The transmittance $T$ and distance $d$ are linked with the expression $T = 10^{-\alpha d/10}$, where $\alpha=0.2$ dB/km is the loss coefficient of the optical fiber. The modulation variance of Alice $V_A$ is adjusted to maintain a signal-to-noise ratio of 0.075 on Bob's side, which allows for a reconciliation efficiency of $\beta=94.8\%$ \cite{JKL+:natphoton13}. The excess noise on Bob's side is $\xi_{\text{Bob}}=0.001$, and the electronic noise of the homodyne detection is $v_{el}=0.01$. For the upper plot, the efficiency of the homodyne detection is assumed to be $\eta=0.6$ while the lower plot corresponds to an efficiency $\eta_{\text{calib}}=0.32$ when taking into account the losses of the optical switch on Bob's signal path.}
     \label{figure:keyrate_countermeasure}
\end{figure}

In Fig. \ref{figure:keyrate_countermeasure}, we compare the theoretical secret key rates against collective attacks \cite{NGA:prl06, GC:prl06} for a CVQKD system that does not implement any countermeasure against the local oscillator calibration attack we proposed and for a system that uses the countermeasure of Fig. \ref{figure:shot_noise_measurement_with_attenuator} with an optical switch on Bob's signal path. In the latter case, the impact of the countermeasure on the secret key rate is twofold. First, the number of pulses that can be used to extract a secret key is diminished by the fraction of pulses chosen at random to compute an estimate of the shot noise; in our numerical analysis, we chose to discard 10\% of the pulses. Second, the efficiency of Bob's measurement apparatus $\eta$ is reduced because of the 2.7 dB losses introduced by the optical switch. For realistic values of all the parameters, we find that the maximum secure distance drops from 80 km to 70 km when implementing this countermeasure.

\section{Conclusion}
We propose a powerful and realistic calibration attack for continuous-variable QKD systems, by which an eavesdropper can make Alice and Bob negotiate a key even for an introduced noise that is above the entanglement breaking limit at which no secret key can be exchanged at any distance. Preventing this attack involves real-time measurement of the shot noise, which is possible but not trivial. Given the relevance of CVQKD technology for high-performance secure communications, this work highlights the importance of rigorously testing the practical security of current implementations.

\section{Acknowledgements}
This research was supported by the French National Research Agency, through the HIPERCOM (2011-CHRI-006) project, by the DIRECCTE Ile-de-France through the QVPN (FEDER-41402) project, and by the European Union through the Q-CERT (FP7-PEOPLE-2009-IAPP) project. P. Jouguet acknowledges support from the ANRT (Agence Nationale de la Recherche et de la Technologie).

\appendix

\section{Local oscillator power measurement and clock signal generation}
Here, we discuss the feasibility of measuring the local oscillator power and generating a trigger signal from the local oscillator without compromising the security of the system.

Reasonable trigger generation functions are of the following form:
\begin{align}
 U_1(t)&=\mathbf{1}_{s(t-r)>x}\\
 U_2(t)&=\mathbf{1}_{s(t-r)-s(t-r-\delta)>0}
\end{align}
The function $U_1$ outputs a positive value at time $t$ if and only if the signal measurement is above the threshold value $x$ at time $t-r$. This corresponds to detecting the beginning of a pulse (when its value is above the threshold $x$) and then delaying the trigger with a chosen delay $r$. The function $U_2$ outputs a positive value at time $t$ if and only if the difference between the signal and the signal delayed of one pulse duration $\delta$ is positive. This presents the advantage of being independent from the signal level but requires to know the pulse duration $\delta$. This cannot be assumed in the context of an active eavesdropper. Both $U_1$ and $U_2$ are of the form $\mathbf{1}_{\phi(s)}$ where $\phi$ is a linear functional of the signal.

Reasonable power measurement functions are of the following form:
\begin{align}
 P=\int_0^{\delta}s(t-s)\alpha^{-s}ds
\end{align}
where $\alpha$ is some nonnegative integration constant. $P$ is a linear form of the local oscillator signal. Since $P$ is not a multiple of $\phi$ for the trigger examples above, there are signals that can be added to the local oscillator signal that do not change the output of $P$ but that change $\phi$. A closer look to this problem shows that it is indeed possible to change $U_i$, $i=1$ or $2$, without changing $P$.

\begin{figure}
\centering
 \includegraphics[width=70mm]{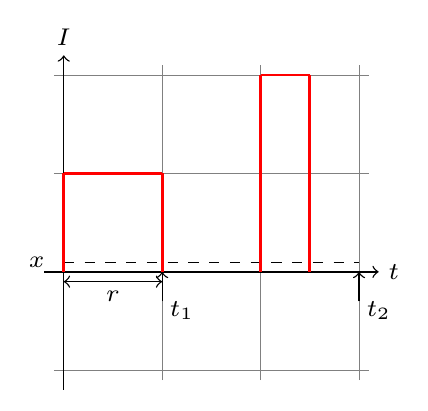}
  \caption{(Color online) This figure shows how two pulses of same energy generate two different trigger signals of rising time $t_1$ and $t_2$.}
   \label{figure:power_vs_trigger}
\end{figure}

A simple example is given in Fig. \ref{figure:power_vs_trigger}. Both local oscillator pulses have the same energy but the rising time of the trigger does not coincide with the end of the pulse.

This analysis shows that, in practice, a calibrated linear relationship between the shot noise level and local oscillator power cannot be used in the presence of an eavesdropper, who will always be able to modify the linear relationship during the QKD run.

\end{document}